\documentclass[12pt]{article}
\usepackage{latexsym}
\usepackage{amssymb}
\usepackage[dvips]{graphicx}
\usepackage{epsfig}
\usepackage{rotating}
\usepackage{amsfonts}
\usepackage{amsmath}

\usepackage{hyperref}
\usepackage{pstricks}

\newtheorem{definition}{Definition}[section]

\newtheorem{lemma}{Lemma}[section]
\newtheorem{corollary}{Corollary}[section]
\newtheorem{theorem}{Theorem}[section]
\newtheorem{example}{Example}[section]
\newtheorem{proposition}{Proposition}[section]

\def\be{\begin{equation}}
\def\ee{\end{equation}}
\def\bea{\begin{eqnarray}}
\def\eea{\end{eqnarray}}

\begin{document}

\begin{titlepage}
\begin{flushright}
ICMPA-MPA/20XX/xx\\
LPT-20XX-xx\\
CPT-P004-2010
\end{flushright}

\vspace{20pt}

\begin{center}

{\Large\bf Linearized Group Field Theory\\
\vspace{10pt}
and Power Counting Theorems}
\vspace{20pt}

Joseph Ben Geloun$^{a,d,e,*}$, Thomas Krajewski$^{a,b,\dag}$,
Jacques Magnen$^{c,\ddag}$\\ and Vincent Rivasseau $^{a,\natural}$

\vspace{15pt}

$^{a}${\sl Laboratoire de Physique Th\'eorique, Universit\'e Paris XI}\\
{\sl CNRS UMR 8627, 91405 Orsay Cedex, France}\\
\vspace{10pt}
$^{b}${\sl on leave, Centre de Physique Th\'eorique, CNRS UMR 6207}\\
{\sl CNRS Luminy, Case 907, 13288 Marseille Cedex 9}\\
\vspace{10pt}
$^{c}${\sl Centre de Physique Th\'eorique, Ecole Polytechnique}\\
{\sl CNRS UMR 7644, 91128 Palaiseau Cedex, France}\\
\vspace{10pt}
$^{d}${\sl International Chair in Mathematical Physics and Applications\\ (ICMPA-UNESCO Chair),
University of Abomey-Calavi,\\
072B.P.50, Cotonou, Rep. of Benin}\\
\vspace{10pt}
$^{e}${\sl D\'epartement de Math\'ematiques et Informatique,\\
Facult\'e des Sciences et Techniques, Universit\'e Cheikh Anta Diop, Senegal}

\vspace{20pt}

E-mail:  $^{*}${\em bengeloun@sun.ac.za},\quad $^{\dag}${\em krajew@cpt.univ-mrs.fr} , \quad
$^{\ddag}${\em magnen@cpht.polytechnique.fr},
\quad $^{\natural}${\em rivass@th.u-psud.fr }

\vspace{10pt}

\begin{abstract}
\noindent
We introduce a linearized version of group field theory.
It can be viewed either as a group field theory over the additive group of a vector space
or as an asymptotic expansion of any group field theory
around the unit group element. We prove exact power counting theorems for any graph of such models.
For linearized {\emph{colored}} models the power counting
of any amplitude is further computed in term of the homology of the graph.
An exact power counting theorem is also established for
a particular class of graphs of the nonlinearized models,
which satisfy a planarity condition.
Examples and connections with previous results
are discussed.
\end{abstract}

\end{center}

\noindent  Pacs numbers:   04.60.-m, 04.60.Pp\\
\noindent  Key words: Group field theory, renormalization, perturbative study.

\end{titlepage}

\setcounter{footnote}{0}

\section{Introduction}
\label{Intro}

Group field theories (GFTs) \cite{boul} are quantum field theories over group manifolds and
can be also viewed as higher rank tensor field theories \cite{Freidel,oriti}.
They provide one of the most promising
framework for a background free theory of quantum gravity in which one sums both
over topologies and geometries. Indeed, each Feynman graph of a $D$ dimensional
GFT can be dually associated with a discrete spacetime ($D$-simplicial complex)
via a specific triangulation
and gluing rules given by the covariance and vertices (basic $D$-simplex) of the theory.
The functional integral formalism then defines a weighted sum over triangulations which
is a weighted ``sum over topologies'' with each weight (amplitude)
related to a ``sum of geometries'' (via a spin foam formalism \cite{rovel}).
Some early results on power counting and non-perturbative resummation of such models
are in \cite{FreiLoua}.

Group field theory is a particular kind of quantum field theory, hence it is natural
to ask the question of its renormalizability. A complete renormalization group analysis for GFTs
is still lacking and the first steps for such a program
have been made only recently \cite{fgo, MNRS, BGMR}.
For the three dimensional Boulatov model
a rigorous power counting theorem has been
obtained for a particular class of graphs called Type I,
 which satisfy a contractibility condition \cite{fgo}.
Given a cutoff $\Lambda$ in the Peter-Weyl field expansion,
Type I graphs diverge as $\Lambda^{B-1}$, where
$B$ denotes the number of ``bubbles''. These bubbles are closed surfaces whose
exact definition is clumsy in general  \cite{fgo}.
 In \cite{MNRS} scaling bounds for Feynman amplitudes of the Boulatov model
have been established together with constructive results. However
in that theory generalized tadpoles dominate over other graphs.

This situation improves markedly in the case of
colored GFT models \cite{gurau}.
Among the nice features of these models we quote:
(1) in the Fermionic case a $SU(D+1)$ global symmetry
in the field indices; (2) a clear definition of ``bubbles'' of any dimension,
which means that the graphs of these colored models
define a full $D$-simplicial complex, not just a 2-complex as in spin foam theory;
(3) a computable homology for this complex implying that
the corresponding dual triangulated objects are manifolds or pseudo-manifolds but with
only point-like singularities;
(4) the absence of generalized tadpoles and of {\emph{tadfaces}}\footnote{A tadpole
is a line which goes twice through the same vertex. A ``tadface'' is a face that goes at least
twice through a line. A generalized tadpole is a subgraph with a single external vertex.};
(5) the existence of an exhaustive sequence of cuts \cite{MNRS} for any graphs of these models.
This leads to better scaling bounds in any dimension for the Feynman amplitudes of these models \cite{BGMR}.

A general power counting theorem  is
a difficult problem even for the colored GFTs. In this paper, we introduce
a further simplification, namely an Abelianized version of GFT
that we shortly call ``linearized GFT''.  This model is also interesting as a kind of
asymptotic limit of GFT at high Fourier frequencies.

We compute the exact power counting for any graph of linearized GFTs. In the colored case
this power counting is reexpressed in terms of the colored homology introduced by Gurau
\cite{gurau}.
The connection with the $\Lambda^{B-1}$ power counting of Type I graphs in
\cite{fgo} is clarified; the linearized colored graphs have power counting $\Lambda^{B-1}$
if and only if their second homology class vanishes.

Exact power counting is also established for amplitudes of the ordinary nonlinarized
models but under a certain planarity condition.

This paper is organized as follows. Section 2 fixes our notations, introduces
the linearized GFTs and establishes the power counting of
their graphs in terms of a certain incidence matrix between lines and faces.
The proof relies on evaluation of determinants of quadratic forms
through their Pfaffian representations in the manner of \cite{gr, RT, KRTW, KRV}
and allows not only to find the exact ultraviolet power counting but also an infrared power counting.
More generally we define the analog of Symanzik polynomials for the linearized models,
which potentially contain much more information. These polynomials are
connected to generalized contraction/deletion moves which for colored graphs 
have been first explored in \cite{gurau2}.

Section 3 contains the basic definitions of colored graphs and their homology
and reexpresses the power counting of linearized colored graphs in term of their additional
structure. We give a computation of the ultraviolet degree of divergence in terms of the
homology of the colored graph as defined in \cite{gurau}.

Section 4 introduces an other technique. The jacket of a graph is defined either
for restricted Boulatov models or for colored models. It is a matrix-like piece of the theory.
An exact non commutative momentum routing is introduced for the dual of the jacket.
When the jacket is planar, the power counting is exactly the number of cycles
of the complement of the jacket.

Finally, Section 5
provides examples, further discussions and conclusions.

\section{Linearized GFTs and their Feynman amplitudes}

An ordinary $D$-dimensional GFT is a field theory for a field
defined over $D$ copies of a group $\tilde{G}$ with a certain gauge invariance
and a certain pattern of interaction. For simplicity, we may consider mostly
the Euclidean 3D case in which $\tilde{G}=SU(2)$ but indicate which of our results are general
for any dimension.
We will also assume that the field is not symmetric under any permutation of the $D$ arguments.
There will also be no symmetrization operators inserted, neither in the propagators nor in the
vertices, so that the category of graphs considered will be the category of stranded graphs
without any twists of the strands.

The Hilbert space of square integrable functions on $\tilde{G}^{D}$ is
$H= L^2 (\tilde{G}^{D})$. The Hilbert space of gauge invariants fields
that is of fields invariant under the ``diagonal'' action
of the group $\tilde{G}$
\begin{equation}
\phi(g_1h,g_2h,\ldots,g_Dh)=\phi(g_1,g_2,\ldots,g_D)
\label{invariances}
\end{equation}
is  a subset $H_0 \subset H$.

The interaction is homogenous of degree $D+1$ and patterned
in the form of a $D$ simplex. For instance in three dimensions this interaction
has the form:
\begin{equation}
S_{int}[\phi]:=\lambda \int\prod_{i=1}^{6}dg_i\ \phi(g_1,g_2,g_3)\phi(g_3,g_4,g_5)\phi(g_5,g_2,g_6)\phi(g_6,g_4,g_1).
\end{equation}
The six integration variables are repeated twice, following the pattern of the edges of a tetrahedron.

In the language of Gaussian measures
the partition function is
\begin{equation}
\label{formal}
Z(\lambda):=\int d\mu_C[\phi]\ e^{-\lambda S_{int}[\phi]},
\end{equation}
where $d\mu_C[\phi]$ is the normalized Gaussian measure whose covariance $C$ can be viewed as
the orthogonal projection from $H$ to $H_0$ implementing  (\ref{invariances})
through an average over the
diagonal action of the group
\begin{equation}
(C\phi)(g_1,g_2,g_3):= \int dh\ \phi(g_{1}h,g_{2}h,g_{3}h).
\end{equation}

Graphs are therefore stranded with $D$ strands per propagator.
The ordinary Feynman rules to build graphs out of propagators and vertices give
for a stranded graph ${\mathcal G}$ of this model an amplitude which is the following formal expression
\begin{equation}
A_{\mathcal G} = \int \prod_e dg_e \prod_\ell dh_\ell  \prod_\ell \prod_{s \in \ell}
\delta(g_{e_s} h_\ell g^{-1}_{e'_s} )
\label{amp1}
\end{equation}
where each line $\ell$ propagates $D$ strands $s$ from edge variable $e_s$ to $e'_s$.

Suppose that we choose both an arbitrary orientation of each line $\ell$ of the graph and of each face
(closed set of strands).
The integration over all edge variables can be explicitly performed, leading  to an other (still
formal) expression for the Feynman amplitude, but solely in terms of an integral over line
variables:

\begin{equation}
 A_{\mathcal G} =
\int
\prod_{\ell \in \mathcal{L}_{{\mathcal G}} } dh_\ell \prod_{  f\in \mathcal{F}_{{\mathcal G}}}
\delta \left({\vec{\prod}_{\ell \in f} h_\ell }\right), \label{amp2}
\end{equation}
where $\mathcal{L}_{{\mathcal G}}$, $\mathcal{F}_{{\mathcal G}}$ are
the set of lines and faces of ${\mathcal G}$, respectively.
The oriented product $\vec{\prod}_{l \in \partial f} h_l$ means that
the variables $h_\ell$ of the delta functions  (which are the $SU(2)$ holonomies along each face)
has to be taken in the following way: $h_\ell$ if the orientations of the line $\ell$
and of the face $f$ coincide and
$h_\ell^{-1}$ if they do not.

Surprisingly the formal amplitude (\ref{amp2}) neither depends of the arbitrary orientation of the
lines, nor of those of the faces! The deep reason is that the two formulas  (\ref{amp1}) and (\ref{amp2}) are equal,
but a more pedestrian way to see this is to exploit carefully the parity of the $\delta$ function
and of the Haar measure under $h \to h^{-1}$.
However a full proof of this fact should address the fact that both (\ref{amp1}) and (\ref{amp2}) are infinite in general.

Indeed multiplication of distributions is not always allowed. Hence formulas (\ref{amp1}) and (\ref{amp2})
suffer from what one should consider from the usual field theoretic point of view as an ultraviolet
\footnote{In the loop quantum gravity context this divergence is often interpreted as an infrared one
for the effective space-time theory, so to underline this ambiguity we might also call it
an {\emph{ultraspin}} divergence.} divergence. However since groups like $SU(2)$ are compact there is no infrared or large volume divergence associated to the group integrations over the $h_\ell$ variables in these
Euclidean GFTs.

A \emph{linearized GFT} is obtained by just changing the group $\tilde G$
to the additive commutative group
of a vector space ${\mathbb R^d}$ and repeating all the arguments above.
It can also be considered as an expansion around the
origin of unity of the group, in which each group element $h_\ell$ is replaced by $e^{i h_\ell}$ and
a Taylor expansion is performed which only retains the linear forms in the new
$h_\ell$ variables in all the $\delta$ functions of (\ref{amp2}). In that
case, $d$ would just be the dimension of the group or of its Lie algebra.
But in the rest of this paper we shall consider $d=1$ for simplicity.

Let us introduce now some terminology.
Non-regular faces, also called tadfaces are faces which goes several times
through the same propagator. A regular
graph is a graph without any tadface. Unfortunately, ordinary GFTs typically generate
tadfaces and non-regular graphs. Non-regular faces play a bit the same
annoying role than tadpoles play in ordinary graphs and ordinary field theory.
They are difficult to capture through an incidence matrix. Although they are somewhat trivial
they may perturb naive versions of contraction-deletion rules.

The linearized GFT amplitude is conveniently defined in terms of
a rectangular $F$ by $L$ incidence matrix
$\varepsilon_{f,\ell}$, where $F= | \mathcal{F}_{{\mathcal G}}|$ is the number of faces of ${\mathcal G}$
and $L= | \mathcal{L}_{{\mathcal G}}|$ is its number of lines.

\begin{definition}
$\varepsilon_{f,\ell}$ is defined as
the algebraic number of passages of the face $f$ through the line $\ell$, counting as $+1$
each passage with coinciding orientation of $f$ and $\ell$ and as $-1$ each passage with
opposite orientation of $f$ and $\ell$. Hence $\varepsilon_{f,\ell} \in {\mathbb Z}$.

For a regular graph $\varepsilon_{f,\ell} \in \{-1, 0, 1\}$ with
\begin{eqnarray}
\varepsilon_{f,\ell}  = \left\{\begin{array}{cc}
 +1, & {\rm if }\;\;\; {\rm or}(l)={\rm or}(f),\;\; \ell \in \partial f  \\
-1,  & {\rm if }\;\;\; {\rm or}(l) = - {\rm or}(f), \;\; \ell \in \partial f\\
0, & {\rm if }\;\;\; \ell \notin \partial f
 \end{array}\right.
\end{eqnarray}
where ${\rm or}(\cdot)$ stands for the orientation.
\label{orient}
\end{definition}

The linearized amplitiude of a regular graph is then given by
\begin{equation}
 A_{\mathcal G} =
\int_{{\mathbb R}^L} \prod_{\ell \in \mathcal{L}_{{\mathcal G}} } dh_\ell \prod_{f \in \mathcal{F}_{{\mathcal G}}}
\delta \left(   \sum_{\ell \in \partial f}   \varepsilon_{f,\ell} h_\ell   \right). \label{amp3}
\end{equation}
Again although the incidence matrix does depend on the chosen orientations,
the amplitude ({\ref{amp3}}) does not depend on them, as we can arbitrarily multiply
each line or column of $\varepsilon_{f,\ell} $ and compensate for that
using either parity of the $\delta$
function or changing $h_\ell$ to $- h_\ell$.

Formula   ({\ref{amp3}}) remains ill-defined and the corresponding amplitude is formally infinite
but now for two distinct reasons:

- the set of $L$ linear forms $\sum_{\ell \in \partial f} \varepsilon_{f,\ell} \, h_\ell $
may typically not be made of independent forms, i.e. the rank $r$ of the
rectangular matrix $\varepsilon_{f,\ell}$ may be smaller than $F$, in which case
the corresponding $\delta$ functions cannot be multiplied. This impossibility
of multiplication of distributions should be considered as
an ultraviolet divergence.

- even if the corresponding multiplication of distributions is regularized, e.g. by replacing
$\delta$ distributions by Gaussian integrals, the rank $r$ of the
rectangular matrix $\varepsilon_{f,\ell}$ could be strictly smaller than $L$, the number of
internal lines. In that case, the corresponding quadratic form in the $h_\ell$ variables
would not be definite, so that the $h_\ell$ integrations do not fully converge at infinity
because there is no decay in some directions in $\mathbb{R}^L$. This kind of divergence
should be considered of infrared type in field theory. It has no analog in the non-linearized
Euclidean group field theories because the group used there is compact; but it has some analogs
in Lorentzian group field theories.

Hence we expect:

\begin{proposition}
The ultraviolet degree of divergence of a linearized regular
GFT graph is $F-r$ and its infrared degree of divergence is $L-r$, where $r$
is the rank of the $\varepsilon_{f,\ell}$ matrix\footnote{Of course this rank is independent
of the chosen orientations to define this matrix.}.
\end{proposition}

To give a precise mathematical content to this statement the simplest an dmost natural proposal, which we follow in this paper, is to regularize
the amplitudes ({\ref{amp3}}) through Gaussian regulation both of the $\delta$ functions
and of the $h_\ell$ integrals.

Hence we are lead to consider infrared regulators $m_\ell$ which break gauge invariance but allow the
amplitude to be well defined even for a noncompact group as is $\mathbb{R}$, and to replace
the $\delta$ function of each face $f$ by a regularized normalized Gaussian function of  the form:
\begin{equation}
\delta_{\Lambda_f} (h) = \frac{\Lambda_f}{\sqrt{2\pi}} \,e^{-\Lambda_f^2\,h^2/2} .
\label{deltc}
\end{equation}

Therefore the ill-defined amplitude of a regular vacuum graph (\ref{amp3}) is re-expressed
using the regularized delta function (\ref{deltc}) as\footnote{Since there is no ambiguity,
we henceforth omit the full notations
 $l\in \mathcal{L}_{{\mathcal G}} $ and $ f\in \mathcal{F}_{{\mathcal G}}$. We also forget inessential multiplicative factors.}
\begin{eqnarray}
 A_{\mathcal G} (m_\ell, \Lambda_f )&=&\int
\prod_{\ell} dh_\ell\, e^{-m^2_\ell h_\ell^2/2}
\prod_{ f } \Lambda_f\,
e^{ - (\Lambda_f^2/2) \left(  \sum_{\ell} \varepsilon_{f,\ell} \, h_\ell   \right)^2}\cr
&=& \int
\left[\prod_{ \ell} dh_\ell
e^{-  m^2_\ell h_\ell^2/2 } \prod_{f} d k_f
e^{ - \Lambda_f^{-2} k_f^2/2   } \right]
e^{ i \sum_{f, \ell } \varepsilon_{f,\ell} h_\ell
 k_f }
\label{ampli2}
\end{eqnarray}
where we forgot inessential normalizations
and we introduced the momentum variables $k_f$ labeled by faces
(see Appendix 1 for useful identities pertaining to that
transformation). The expression (\ref{ampli2}) corresponds to a kind of phase
space representation of the amplitude.

The limit ({\ref{amp3}}) is formally recovered as all $\Lambda_f \to \infty$ and all $m_\ell \to 0$.

We start now the derivations of our main results
on exact power counting theorems. At first we will discuss
the case of vacuum graphs and then we will focus on graphs
with external legs. The proof of the second
theorem will be shortened following similar steps as for
the situation without external legs.

\begin{theorem}\label{theo1}
For any connected vacuum graph $\mathcal{G}$ of the linearized GFT,
we have
\begin{equation}
A_{\mathcal{G}} (1, \Lambda) \simeq_{\Lambda \to \infty}  K \Lambda^{F- r}.
\label{powuv}
\end{equation}
where $K$ is an inessential factor. Similarly
\begin{equation}
A_\mathcal{G} (m, 1) \simeq_{m \to 0}  K m^{-(L-r)}.
\end{equation}
\end{theorem}

\noindent {\bf Proof.}
We shall deduce all these results from a more general formula which computes
the Feynman amplitudes  $A_{\mathcal G} (m_\ell, \Lambda_f )$ in terms of a topological
``Symanzik polynomial".

We introduce the vector $\mathfrak{h}= (h_1,h_2,\ldots, h_L, k_1,k_2,\ldots, k_F)^t$
and the matrix defined by its block sub-matrices
\begin{eqnarray}
\mathfrak{M} = \left(\begin{array}{cc}
m^2_\ell\,\delta_{\ell \ell'}& i\, ^t\varepsilon_{\ell, f'}\\ &\\
i\, \varepsilon_{f, \ell'} & \Lambda_f^{-2} \delta_{ff'}
 \end{array}\right)
\label{matfram}
\end{eqnarray}
Gaussian integration gives, up to inessential normalizations
\begin{equation}
  A_{\mathcal G} =\int
\left[\prod_{\ell} dh_\ell\,\prod_{f} d k_f\right]
e^{-\mathfrak{h}^t\,  \mathfrak{M} \,   \mathfrak{h} /2 } =
  \frac{1}{\sqrt{{\rm Det} \, \mathfrak{M} }}.
\end{equation}
There exists many ways to determine the behaviour of
$A_{\mathcal G}$ at large $\Lambda_f$'s. For instance, the Schur
complement formula gives (\ref{powuv}), for  $m_\ell =1 \;\; \forall \ell $,
$\Lambda_f = \Lambda \;\;  \forall f$.
However this method does not give the detailed dependence
of $A_{\mathcal G} (m_\ell, \Lambda_f )$
in all joint variables. More information is obtained by the method of \cite{gr}
which provides a Pfaffian expansion of a determinant
$\det (D+R)$,
where $D$ is diagonal and $R$ is antisymmetric.
It expands the determinant as:
\begin{eqnarray}
{\rm Det}\,(D+R)= \sum_{I \subset{1,2,\ldots,N}}
\prod_{i\in I} D_{ii}\; {\rm Pf}^2\, R_I,
\end{eqnarray}
where $R_I$ is the matrix obtained from $R$ after removing
lines and columns of indices in $I$.

We first remark that the determinant of $\mathfrak{M}$
coincides with the determinant of
\begin{eqnarray}
M = \left(\begin{array}{cc}
m^2_\ell\delta_{\ell\ell'}& - ^t\varepsilon_{\ell, f'}\\ &\\
\varepsilon_{f, \ell'} & \Lambda_f^{-2} \delta_{ff'}
 \end{array}\right)
\end{eqnarray}
obtained from $\mathfrak{M}$ after multiplying the block $i\varepsilon^t$
by $i$ and the block $i\varepsilon$ by $-i$. This decomposes $M$ as the sum of a diagonal and
an antisymmetric matrix. Hence
\begin{eqnarray}
{\rm Det} \, \mathfrak{M} =
 \sum_{\stackrel { I \subset \{1,2,\dots, L \}}{
 J \subset \{1, 2 , \ldots F \}}}
\prod_{\ell \in I } m^2_\ell \prod_{f \in J }  \Lambda_f^{-2}  \; {\rm Pf}^2 \varepsilon_{\hat I, \hat J}
\label{dete}
\end{eqnarray}
where ${\rm Pf} \varepsilon_{\hat I, \hat J}  $ means the
Pfaffian of $\varepsilon$ with lines and columns in $I $ and $ J $ erased.

Remark that this is a polynomial in $m^2_\ell$ and $\Lambda_f^{-2}$ with positive integer coefficients
which we could call the Symanzik topological polynomial of the graph. It should
obey deletion-contraction relations.

We are now in position to find the scaling behaviour
of $A_{\mathcal G} \simeq ({\rm Det} \,\mathfrak{M})^{-1/2}$.
This can be done by putting $m_\ell =1 \;\; \forall \ell $, $\Lambda_f = \Lambda \;\;  \forall f$ and
identifying the non-zero monomial with the lowest power
$p$ in $\Lambda^{-2}$ in (\ref{dete}).
One notices that because equation (\ref{dete})  is a sum of squares
this is equivalent to find the subset $J$ with the minimal
order such that ${\rm Pf} \varepsilon_{\hat I, \hat J}  \neq 0$, in other words to
find the minor of maximal rank in $\varepsilon$. Then $p = \vert J \vert = F -{\rm rank}\,\varepsilon$,
since the rank of $\varepsilon$ is the cardinal of $\hat J $, the complement of $J$ in $\{1, 2, \ldots F\}$.
We are led to $A_{\mathcal G} \simeq  ( \Lambda^{-2 p})^{-1/2}=  \Lambda^p  =
\Lambda^{F-{\rm rank}\,\varepsilon}$. This achieves the proof of the ultraviolet power counting.
Similarly to get the infrared power counting one should
identifying the non-zero monomial with the lowest power
$q$ in $m^{2}$ in (\ref{dete}), hence find the subset $I$ with the minimal
order such that ${\rm Pf} \varepsilon_{\hat I, \hat J}  \neq 0$. This means to
find again the minor of maximal rank in $\varepsilon$. Therefore $q = \vert I \vert = L -{\rm rank}\,\varepsilon$,
since the rank of $\varepsilon$ is the cardinal of $\hat I $, the complement of $I$ in $\{1, 2, \ldots L\}$.
\hfill
$\square$

We consider now  graphs with external legs.
Such a graph has $L$ internal lines of index $\ell $, $N$ external
legs of indices $j$, $F$ internal (closed) faces of indices $f$and
$F'=DN/2$ open strands of indices $s$. We define a rectangular $L+N$ by $F+F'$ incidence matrix
by choosing again an arbitrary orientation of each line whether it is internal or external and
each face or open strand.

An interesting quantity which corresponds naturally to Feynman amplitudes with fixed external momenta
is (up to inessential normalization constants)

\begin{eqnarray}
 A_{\mathcal G} (k_1, \ldots, k_s, \ldots, k_{F'}) &=&
\prod_j m_j^{-1} e^{ -  m_j^{-2} (\sum_s \varepsilon_{j,s} k_s)^2 }
\prod_{s} e^{ - \Lambda_s^{-2} k_s^2 }
\tilde A_{\mathcal G} (k_1, \ldots, k_s, \ldots, k_{F'}) \cr
\tilde A_{\mathcal G} (k_1, \ldots, k_s, \ldots, k_{F'}) &=&
\int
\prod_{\ell } dh_\ell\, e^{- m_\ell h_\ell^2/2} \prod_{ f } \Lambda_f\,
e^{ - \Lambda_f^2 \left(  \sum_{\ell} \varepsilon_{f,\ell} \, h_\ell   \right)^2}
\prod_{ s }
e^{ i (\sum_\ell \varepsilon_{\ell, s } \, h_\ell  k_s  )}
\cr
= \int
\prod_{ l } dh_l
e^{-  m^2_\ell h_\ell^2/2 } && \prod_{f} d k_f
e^{ -   \Lambda_f^{-2}  k_f^2 /2}  \;
e^{ i \sum_{\ell, f } \varepsilon_{f,\ell} h_\ell k_f+ i \sum_{\ell, s} \varepsilon_{s,\ell}h_\ell k_s }
\cr
&=& ({\rm Det} M) ^{-1/2}  \exp { [-\;   ^tK \,^tS\, M^{-1}\, S K ] }
\label{ampli3}
\end{eqnarray}
where $M = D + R$, R being the antisymmetric $L+F$ by $L+F$ matrix
whose only non zero upper triangular part is the $L$ by $F$ rectangular matrix $\varepsilon_{\ell, f}$
and $S$ being the $L+F$ by $DN/2$  matrix with only non trivial part  $\varepsilon_{s,\ell}$.
$K$ is the vector with components $k_s$.

The quadratic form $^tS M^{-1} S $ is a rational fraction in the $\Lambda_f^{-2}$ and $m^2_\ell$ variables
whose denominator is an analog of the first Symanzik polynomial and whose numerator
is an analog of the second Symanzik polynomial.

At zero external momenta the scaling properties are the same
as those of  Theorem \ref{theo1} but for the graph with the reduced set of internal faces
and internal lines. This graph has typically no longer $D$ strands per line but less, since the external strands have been deleted.

\section{Colored GFTs and their linearization}

\subsection{Colored GFTs}

Colored GFTs have many improvements over ordinary GFTs. All their graphs are regular and equipped with
canonical orientations of lines, faces, and higher bubbles. The graphs have a well defined homology \cite{gurau} and
admit an exhaustive sequence of cuts \cite{MNRS}.
We show in this section that the  power counting
of any corresponding linearized colored amplitudes, still expressed through Theorem \ref{theo1},
can be also computed in terms of their homology.

Let us first review the definition of the
colored models along the lines of \cite{gurau}.

Colored GFT models in dimension $D$ are defined in terms of $D+1$ complex Bosonic or Fermionic fields
denoted  by $\phi^\ell$, where $\ell=1,2,\ldots,D +1$ is referred to as the color index or simply color. These
fields are also defined on $\tilde G^D$, where e.g. $\tilde G = SU(2)$.
The dynamics is described by an action expressed  in term of left-invariant fields as
\begin{eqnarray}
 &&S^{D}[\phi] :=   \int \,  \prod_{i=1}^D\,dg_i \,e^{-g_i^2/2} \;\;\;
 \sum_{\ell=1}^{D+1}\, {\bar\phi}^\ell_{1,2,\ldots,D}\;\;
 \phi^\ell _{D,\ldots,2,1} \crcr
&&+ \lambda_1
\int\prod dg_{i^j}\,e^{-g_{i^j}^2/2}
\phi^1_{1^2,1^3,\ldots, 1^{(D+1)}}\; \phi^{(D+1)}_{(D+1)^1,(D+1)^2,(D+1)^3,\ldots,(D+1)^{D}}\;
 \phi^{D}_{D^{D+1},D^{1},D^2,\ldots, D^{D-1}}\ldots\cr
&&\qquad \ldots  \ldots \phi^3_{3^{4},3^{5},\ldots,3^{D+1}, 3^{1},3^2}\;
 \phi^2_{2^{3},2^4,\ldots,2^{D+1}, 2^{1}} \prod_{j\ne i}^{D+1} \delta(g_{i^j}(g_{j^i})^{-1})\cr
\crcr
&&+ \lambda_2
\int\prod dg_{i^j}\,e^{-g_{i^j}^2/2}\;
\bar\phi^1_{1^2,1^3,\ldots, 1^{(D+1)}}\; \bar\phi^{D+1}_{(D+1)^1,(D+1)^2,(D+1)^3,\ldots,(D+1)^{D}}\;
\bar \phi^{D}_{D^{D+1},D^{1},D^2,\ldots, D^{D-1}}\ldots\cr
&&\qquad \ldots  \ldots \bar\phi^3_{3^{4},3^{5},\ldots,3^{D+1}, 3^{1},3^2}\;
 \bar\phi^2_{2^{3},2^4,\ldots,2^{D+1}, 2^{1}} \prod_{j\ne i}^{D+1} \delta(g_{i^j}(g_{j^i})^{-1})\cr
&& \label{action}
\end{eqnarray}
where ${\bar\phi}^\ell \phi^\ell $ are quadratic mass terms
and integrations are performed over copies of $\tilde G$ using products
Haar measures, and $\lambda_{1,2}$ are the coupling
constants. We have introduced the shorthand notation
$\phi^\ell(g_{\ell^i},g_{\ell^j},\ldots, g_{\ell^k})=\phi^\ell_{\ell^i,\ell^j,\ldots, \ell^k}$,
where the label of group element $g_{\ell^k}$ reminds us the link of two
colors $\ell$ and $k$  in the vertex. The left-invariant constraint is again
given by (\ref{invariances}).

Again the linearized version of this model
is simply expressed by replacing the group $\tilde G$ by ${\mathbb R}$.

\begin{figure}
\centering
\centering
\includegraphics[angle=0, width=14cm, height=6cm]{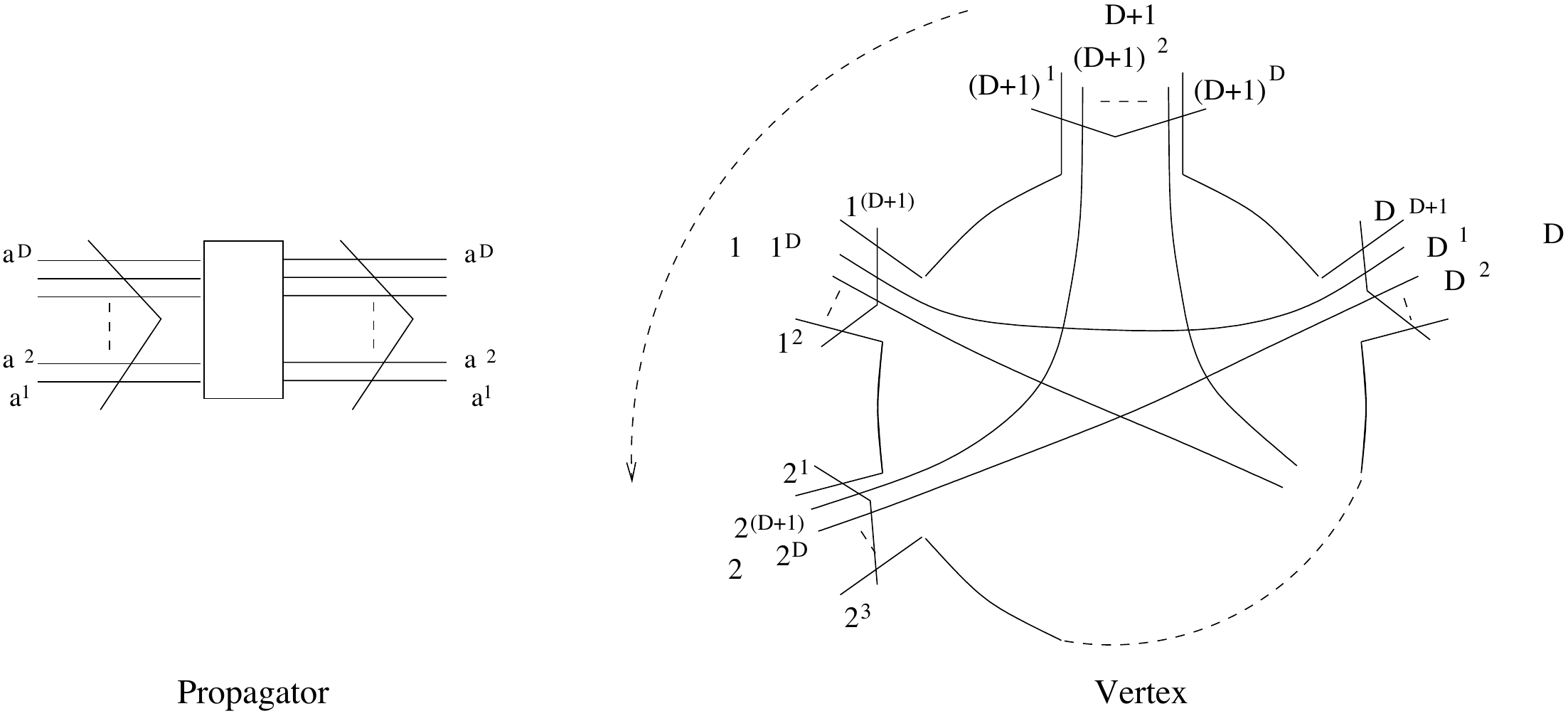}
\caption{
{\small Propagator and vertex $\phi^{D+1}$ in $D$ dimensional GFT. }}\label{colvertprop}
\end{figure}

The propagator and vertex $\phi^{D+1}$ are drawn in Fig. \ref{colvertprop};
the vertex for $\bar \phi^{D+1}$ can be easily found by reversing the order of
all labels in Fig. \ref{colvertprop}.

An interesting property of colored GFTs is that the coloring can be used to define
canonical orientations and a full homology for any colored graph, including
external legs \cite{gurau,gurau2}. More precisely
it is shown in \cite{gurau,gurau2} that the $p$-bubbles and boundary $p$-bubbles define
two cellular complexes $(\mathfrak{U}^*, d_*)$ and $(\mathfrak{U}^*_\partial, d^\partial_*)$
and therefore two cellular homologies induced by two boundary operators
$d_*$ and $d_*^\partial$ such that
\begin{equation}
d_p :\, \mathfrak{U}^p \to \mathfrak{U}^{p-1}\,,\qquad
d_p^\partial :\, \mathfrak{U}^p_\partial \to \mathfrak{U}^{p-1}_\partial.
\end{equation}
These cellular complexes are made of $k$-bubbles, which are connected components of
strands of subsets of $k$-colors.
Note that the sets of such 0-bubbles $\mathfrak{B}^{0}$,
of 1-bubbles $\mathfrak{B}^{1}$ and of 2-bubbles $\mathfrak{B}^{2}$
naturally coincide with the sets of vertices ${\mathcal V}$, of lines ${\mathcal L}$
and of faces ${\mathcal F}$, respectively.

In order to define properly
the simplicial complex of the boundary, one needs some further ingredients by making as a
connected object the boundary which can be {\it a priori} disconnected.
This is realized by the ``Gurau pinching'' \cite{gurau2}, introducing new vertices
(any boundary $D$-simplex itself can be dually associated with a vertex $\phi^D$ for a $D$ dimensional group field model) in the theory which mainly renders connected the dual boundary graph.

Then one can show that indeed $d_p \circ d_{p+1} =0$ and $d^\partial_p \circ d^\partial_{p+1} =0$.
We also define the extreme differential operators $d_{D+1}:=0$ and $d^\partial_{D}:=0$.
From this stage, the homology groups can be defined either for
$(\mathfrak{U}^*, d_*)$ or for $(\mathfrak{U}^*_\partial, d^\partial_*)$.
We have for the ``vacuum'' complex $(\mathfrak{U}^*, d_*)$ the
$p$-th homology group defined by $H_p = {\rm Ker}\,d_p / {\rm Im}\,d_{p+1}$
and for the boundary complex $(\mathfrak{U}^*_\partial, d^\partial_*)$,
$H^\partial_p = {\rm Ker}\,d^\partial_p / {\rm Im}\,d^\partial_{p+1}$
which can be called the $p$-th boundary homology group.

For a general graph  $\mathcal{G}$ with external legs, one identifies
easily the boundary graph $\mathcal{G}_\partial$ generated by external
strands pinched and a remaining part denoted by
$\bar{\mathcal{G}}=\mathcal{G}\setminus\mathcal{G}_\partial$
which is generally a sum of vacuum
graphs. Using the definitions
of cellular complexes for vacuum graphs
$(\mathfrak{U}^*,d_*)$ and for a boundary graph $(\mathfrak{U}^*_\partial,d_*^\partial)$,
one can define the complex $(\mathfrak{V}^*,\delta_*)$ for a graph $\mathcal{G}$ with external legs
such that
\begin{eqnarray}
\mathfrak{V}^p = \mathfrak{U}^p \oplus \mathfrak{U}^p_\partial, \quad \forall p,\\
\delta_p = d_p \oplus d_p^\partial, \quad \forall p.
\end{eqnarray}
A quick checking proves that $\delta_{p}\circ \delta_{p+1}= (d_p\circ d_{p+1})
\oplus (d_p^\partial\circ d^\partial_{p+1})=0$, since there is no cross terms.
The definition of homology group of graphs follows naturally
{\boldmath$H$}$_p= {\rm Ker}\,\delta_p /  {\rm Im}\,\delta_{p+1}$.

As a consequence of the direct sum, we have
 ${\rm Ker}\,\delta_* =  {\rm Ker}\,d_{*}\oplus  {\rm Ker}\,d^\partial_{*}$,
 ${\rm Im}\,\delta_{*}=  {\rm Im}\,d_{*}\oplus  {\rm Im}\,d^\partial_{*}$
 and {\boldmath$H$}$_* \equiv H_{*}\oplus H^\partial_{*}$.
Finally it can be proved that
\begin{eqnarray}
&&{\rm Ker}\,d_{D} = \mathbb{Z},\;\;\;\;{\rm Im}\,d_{D}= \oplus_{k = 1}^{B^D-1}\;\mathbb{Z},\;\;\;\;
H_D \equiv  {\rm Ker}\,d_{D}; \label{holt}\\&&\cr
&&{\rm Ker}\,d^\partial_{D-1} = \mathbb{Z},\;\;\;\;{\rm Im}\,d^\partial_{D-1}= \oplus_{k = 1}^{B_\partial^{D-1}-1}\;\mathbb{Z},\;\;\;\;
H^\partial_{D-1} \equiv  {\rm Ker}\,d^\partial_{D-1}.
\label{holb}
\end{eqnarray}

\subsection{Homology formula for linearized color power counting}

We can now define {\it canonically } the incidence matrix $\varepsilon_{f,\ell}$,
using the ordering of the colors.
\begin{definition}
\begin{eqnarray}
\varepsilon_{f,l}  = \left\{\begin{array}{cc}
 +1, & {\rm if }\;\;\; {\rm or}(l)={\rm or}(f),\;\; l \in \partial f  \\
-1,  & {\rm if }\;\;\; {\rm or}(l) \neq {\rm or}(f), \;\; l \in \partial f\\
0, & {\rm if }\;\;\; l \notin \partial f
 \end{array}\right.
\end{eqnarray}
where ${\rm or}(\cdot)$ stands for the orientation, and

- any line is oriented from $\bar \phi$ to $\phi$,

- any face is oriented \cite{gurau} according to its lines with highest color index.
Hence each face being bicolored, for such a face $f$ made of the two alternating colors $i_1 < i_2$
we have $\varepsilon_{f,i_2} =+1$ and $\varepsilon_{f,i_1} =-1$.
\label{orient1}
\end{definition}
Then the amplitude of a colored graph ${\mathcal G}$ in the linearized model
is formally given by
\begin{equation}
 A_{\mathcal G} =
\int \prod_{ \ell \in \mathcal{L}_{{\mathcal G}} } dh_l\,
\prod_{f \in \mathcal{F}_{{\mathcal G}}}
\delta \left(\sum_{\ell  \in \partial f} \varepsilon_{f,\ell } \, h_\ell \right).
\label{ampli}
\end{equation}
Let us now relate the power counting of Theorem \ref{theo1} to a homology formula
\begin{theorem}
\label{theo2}
For any connected colored vacuum graph $\mathcal{G}$ of the linearized
$D$ dimensional colored group field model, the amplitude of $\mathcal{G}$ behaves as
\begin{equation}
 \mathcal{A}_\mathcal{G} \simeq K
 \Lambda^{\sum_{k=3}^{D+1} (-1)^{k-1} B^{k} + \sum_{k=2}^{D-1} (-1)^{k} h_k}, \;\;\;
          \;\;B^{D+1}:=1
\end{equation}
where $K$ is an inessential factor,
$B^k$ is the number of $k$-bubbles and  $h_k$ stand for Betti numbers for $\mathcal{G}$
namely ${\rm dim}H_k$.
\end{theorem}
\begin{corollary}
\label{coroll1}
In three dimensions,
any connected colored vacuum linearized amplitude $\mathcal{G}$ behaves as
\begin{equation}
 \mathcal{A}_\mathcal{G} \simeq K
 \Lambda^{B^3 -1 + h_2 }
\end{equation}
where $h_2$ is the second Betti number.
\end{corollary}
Hence we recover the same formula than for Type I graphs in \cite{fgo}, except for the
correcting factor $h_2$, which must therefore vanish for Type I graphs.

The proof of Theorem \ref{theo2} simply  follows from Theorem \ref{theo1} and
the following lemma
\begin{lemma}
Let $\mathcal {G}$ be a colored vacuum graph in dimension $D\geq 3$ \
with $F$ faces and an incidence matrix $\varepsilon$
between its faces and lines, then
\begin{equation}
F- {\rm rank}\, \varepsilon =
\sum_{k=3}^{D+1} (-1)^{k-1} B^{k} + \sum_{k=2}^{D-1} (-1)^{k} h_k, \;\;\;
          \;\;B^{D+1}:=1
\end{equation}
where $B^k$ is the number of $k$-bubbles
and  $h_k$ the $k$-th Betti number of $\mathcal{G}$.
\end{lemma}
{\bf Proof.} This is a straightforward derivation after noting
the fact that, by our definition (\ref{orient1}), the matrix $\varepsilon$ coincides with
the matrix of the differential operator $d_2$ defined in \cite{gurau}.
Introducing the differential complex
defined for a fixed graph $\mathcal{G}$
\begin{equation}
0 \;\stackrel{d_0}{\leftarrow}\; \mathfrak{U}^{0}\;
 \stackrel{d_1}{\leftarrow}\;
\mathfrak{U}^{1}\;
 \stackrel{d_2}{\leftarrow}\;
\mathfrak{U}^{2}\;
 \stackrel{d_3}{\leftarrow}\;
\mathfrak{U}^{3}\;
\leftarrow \ldots\ldots\;
 \stackrel{d_{D-1}}{\leftarrow}\;
\mathfrak{U}^{D-1}\;
 \stackrel{d_D}{\leftarrow}\;
\mathfrak{U}^{D}\;
\stackrel{d_{D+1}}{\leftarrow}\;0
\end{equation}
we have by a simple recurrence
\begin{eqnarray}
F- {\rm rank}\, \varepsilon &=&
 B^{2}- {\rm dim} \,{\rm Im}\, d_2
= {\rm dim} \,{\rm Ker}\, d_2  =
{\rm dim} \,{\rm Im}\, d_3 + {\rm dim} \,H_2\cr
&=&   B^{3}-{\rm dim} \,{\rm Ker}\, d_3+{\rm dim} \,H_2 =
 B^{3} - {\rm dim} \,{\rm Im}\, d_4 - {\rm dim} \,H_3 +  {\rm dim} \,H_2
=\ldots \cr
&=&
\sum_{k=3}^{D-1} (-1)^{k-1} B^{k} + (-1)^{D-1}
               {\rm dim} \,{\rm Ker}\, d_{D-1}
+ \sum_{k=2}^{D-2} (-1)^{k} {\rm dim} \,H_k   \cr
&=&
\sum_{k=3}^{D-1} (-1)^{k-1} B^{k} + (-1)^{D-1}
              ({\rm dim} \,{\rm Im}\, d_{D} + {\rm dim} \,H_{D-1})\cr
&&+ \sum_{k=2}^{D-2} (-1)^{k} {\rm dim} \,H_k.
\end{eqnarray}
The result follows after noting that from (\ref{holt}), ${\rm dim}\, {\rm Im}_{D} = B^D-1$.
\hfill $\square$

Similarly
\begin{theorem}
\label{theo3}
For any connected colored graph $\mathcal{G}$ with $N$ external legs of the linearized $D$ dimensional
colored group field model, the amplitude of $\mathcal{G}$ at fixed bounded external momenta behaves as
\begin{equation}
 \mathcal{A}_\mathcal{G} \simeq K
 \Lambda^{\sum_{k=3}^{D+1} (-1)^{k-1} B^{k} + \sum_{k=2}^{D-1} (-1)^{k} h_k}, \;\;\;
          \;\;B^{D+1}:=1
\end{equation}
where $K$ is an inessential factor, $B^k$ is the number of $k$-bubbles
and  $h_k$ stand for the Betti numbers for $\bar {\mathcal{G}}$,
the graph obtained by deleting the external open strands in ${\mathcal{G}}$.
\end{theorem}
%
%

\section{Power counting for graphs with a planarity condition}

 Here  we present a comparison of the power counting for a particular
 class of graphs  in the linear and non-linear model.
Most of our results in this section
are valid for the graphs of the $BF$ theory with group $G$  in  dimensions $D=2,3,4$,
be it colored or not.  It turns out that the amplitude
is largely independent of the group $G$ for a specified category of graphs.

First, recall that the lines GFT  graphs in these dimensions are made of $D$ strands.
Their vertices of valence $D+1$ involve a specific routing of the incoming strands.
 Closed circuits followed by the strands define the faces and for a vacuum diagram,
 the amplitude reads
 \begin{equation}
 {\cal A}_{{\mathcal G}}=\int \prod_{l} dh_{l}\,\prod_{f}\delta_{\Lambda}\bigg(\vec{\prod}_{l\in \partial f}h_{l}\bigg).\label{amplitude}
 \end{equation}
 The integration measure on the group is assumed to be normalized,
for instance by taking the normalized Haar measure on a compact Lie group or the
 Lebesgue measure with a Gaussian weight for ${\mathbb R}$.
The delta functions are regularized either using the heat-kernel
 \begin{equation}
 \delta_{\Lambda}(h)=\sum_{\rho \;\;
 \mbox{\tiny irreps}}d_{\rho}\mbox{Tr}_{\rho}(h)\,e^{-\Lambda ^{-2}c_{\rho}},
 \end{equation}
 for $G$ a simple Lie group and
 \begin{equation}
 \delta_{\Lambda}(h)=\frac{\Lambda}{(2\pi)^{1/2}}\,e^{-\Lambda^{-2}h^{2}/2},
 \end{equation}
 for $G={\mathbb R}$, so that we recover the Dirac distribution on $G$ in the limit $\Lambda\rightarrow \infty$.
 The heat kernel regularization has the obvious advantage of preserving the positivity but any other
regularization fulfilling $\delta_{\Lambda}(gh)=\delta_{\Lambda}(hg)$ could also be used in the following argument.

\begin{definition}
For a GFT graph ${\mathcal G}$, we define its jacket $\overline{{\mathcal G}}$ as
the ribbon graph made of the 2 external strands.
\end{definition}

Note that the jacket can be defined only because we considered only models generating
 untwisted stranded graphs without any symmetrization on the strands. The jacket then
fully determines the underlying GFT graph. For a colored model the
 jacket can be uniquely defined in terms of the colors only instead of the external strands.
This definition is motivated by the following result, which exhibits a class of graphs for which
the amplitude do not depend on the nature of the group $G$.

\begin{theorem}
For a connected GFT vacuum graph ${\mathcal G}$ and whose jacket is planar, the amplitude reads
 \begin{equation}
{\cal A}_{{\mathcal G}}= \big[\delta_{\Lambda}(1)\big]^{N+1},
\end{equation}
where $N$ is the number of circuits followed by the $D-2$ strands that do not define its jacket.
\end{theorem}
\noindent{\bf Proof}:
It is convenient to pass to the dual $\overline{{\mathcal G}}^{\ast}$. In this case the faces of
$\overline{\mathcal G}$ become vertices of $\overline{{\mathcal G}}^{\ast}$ and the Dirac distribution
around faces become more familiar momentum conservation around vertices. Lines of $\overline{{\mathcal G}}$
are in bijection with lines of $\overline{{\mathcal G}}^{\ast}$, which may be drawn perpendicularly to
those of $\overline{\mathcal G}$. Note however that momenta are group elements and do not necessarily commute.
This is taken into account by the cyclic ordering at the vertices of $\overline{\mathcal G}^{\ast}$.
Next, choose a spanning tree ${\cal T}$ in $\overline{\mathcal G}^{\ast}$.
Because of momentum conservation, momenta on the lines of ${\cal T}$ are entirely determined
by the momenta in $\overline{{\mathcal G}}^{\ast}-{\cal T}$.  Then, the product around the faces of
the $\delta$-function  pertaining to the faces of $\overline{{\mathcal G}}$ in \eqref{amplitude}
can be written as a product of these $\delta$-functions on the lines of ${\cal T}$ times a $\delta$-function
expressing the momentum conservation around ${\cal T}$. More precisely, let us choose a vertex on ${\cal T}$,
the root, and orient all lines  ${\cal T}$ from the root to the leaves. The lines not in ${\cal T}$ join
two vertices in ${\cal T}$ and we orient them towards the root, that is, in such a way that they form a
consistently oriented cycle with the lines of ${\cal T}$ between these two vertices.
Finally, all faces in $\overline{\mathcal G}$  inherit the orientation of the surface it is embedded in.
We denote by ${\cal T}|_{l}$ the tree formed by the lines of ${\cal T}$ between the root and $l$, $l$ not included.
Then,
\begin{equation}
\prod_{f\,\mathrm{faces\,of }\, \overline{{\mathcal G}} }\delta_{\Lambda}\bigg(\vec{\prod}_{l\in \partial f}h_{l}\bigg)
=
\delta_{\Lambda}(h_{{\cal T}})\prod_{l\in{\cal T}}\delta_{\Lambda}({h}_{{\cal T}|_{l}})\label{deltajacket}
\end{equation}
where $h_{\cal T'}$ is product of the cyclically ordered momenta incoming to the tree ${\cal T'}$.

\begin{figure}\label{dual}
\centering
\centering
\includegraphics[angle=0, width=6.5cm, height=5cm]{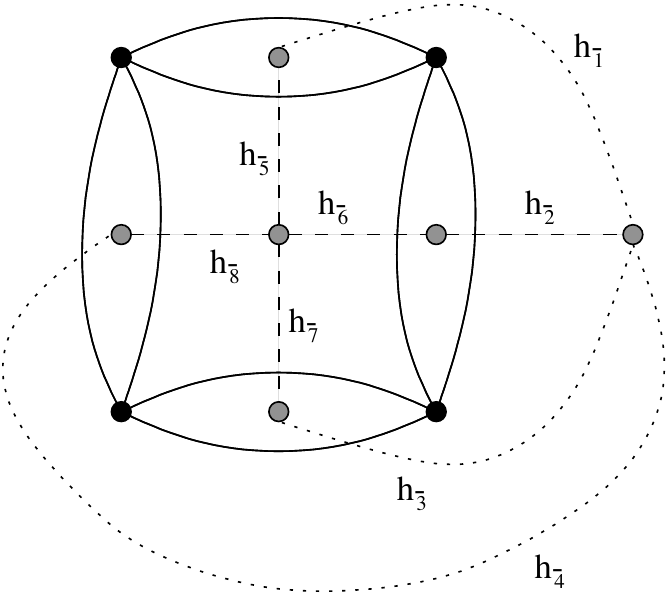}
\caption{
{\small A graph, its dual and a spanning tree on the dual. }}
\end{figure}

If we further assume that $\overline{{\mathcal G}}$ is planar, then the lines $l\notin{\cal T}$ incident to ${\cal T}$
are either nested or disjoint, so that each line appears at most once in each $\delta$ functions in \eqref{deltajacket}.
Accordingly, $\delta_{\Lambda}(h_{{\cal T}})=\delta_{\Lambda}(1)$.
Consider now any closed circuit followed by the $D-2$ strands that do not form the jacket and draw
it on $\overline{{\mathcal G}}^{\ast}$ using perpendicular lines.
This circuit encircles a certain number
of vertices of $\overline{{\mathcal G}}^{\ast}$ and momentum conservation implies that the product
of the $h_{l}$  along the circuit is 1. For instance, on the example of figure \eqref{dual}.
the tree momenta $h_{2},h_{5},h_{6},h_{7},h_{8}$ are determined in terms
of the loop momenta $h_{1},h_{3},h_{4}$

\hfill $\Box$\\

In the general case of a non planar jacket, \eqref{deltajacket} still holds.
 However, the lines not in ${\cal T}$ may have crossings, so that
we have factors of the type $\delta_{\Lambda}(h\cdots h^{-1}\cdots)$, with the dots denoting group elements other than $h$.
In particular $\delta_{\Lambda}(h_{\cal T})$ encodes the defining relation of the fundamental group of the surface
$\overline{{\mathcal G}}$ is drawn on. Then, the amplitude cannot be computed without using specific
group theoretical relations, unless the $\delta$-functions of the closed circuits yield some extra simplifications.
Nevertheless, troublesome factors of the type  $\delta_{\Lambda}(h\cdots h^{-1}\cdots)$ are trivial
for the commutative group ${\mathbb R}$. Since the divergence of the amplitude is a consequence of the
redundancy of the $\delta$ functions,
 it may be reasonably asserted that if a diagram is divergent in the non-linear theory,
so it is in the linearized one. \hfill $\Box$\\

\begin{example}[The octahedron]
Consider a graph whose structure is an octahedron (6 vertices and 12 lines). Its jacket is planar and the strands
not in the jacket form 3 circuits. This immediately yields
\begin{equation}
{\cal A}_{\mbox{\tiny octahedron}}= \big[\delta_{\Lambda}(1)\big]^{4}.
\end{equation}
\end{example}





\section{Conclusion and Discussions}

We have defined the linearized group field theories and established 
their power counting.
The result sheds more light on a result of \cite{fgo}, where the power counting for a particular class of graphsin three dimensions is given
in terms of the number of their 3-bubbles.

Our results show how this formula has to be completed in the linear and colored situation
by the alternating sum of the number of higher dimensional bubbles
and Betti numbers for the graphs. This study also shows the simplicity of
colored models with their more regular structures.

This work has introduced
also other power counting results under planarity conditions.

Apart from these properties,
we remark that for linearized colored graphs the power counting
is $B^3-1$, like for Type I graphs, if and only if $h_2 =0$.
This proves that
\begin{theorem}
Every Type I graph in the sense of \cite{fgo} must have $h_2 =0$.
\end{theorem}

It is however not so easy to build colored graphs with $h_2 \ne 0$.
Remark that the three colored graphs considered in \cite{ gurau} all have $h_2 =0$.
However such graphs should exist. By Poincar\'e duality, any compact
manifold (such as the 3-torus $T_3$) for which one has $h_1  \ne 0$ has
also $h_2 = h_1 \ne 0$. From a theorem by Kaufmann, any piecewise-linear manifold admits a colored
triangulations \cite{Lins}, so there should exist such graphs but we have not found any explicit example yet.

We should explore elsewhere the properties  of the
Symanzik polynomials for tensor theories introduced
in this paper under deletion-contraction of lines, faces or higher ``bubbles"
and their link to the polynomials introduced in \cite{gurau2}.

\begin{figure}
\centering
\centering
\includegraphics[angle=0, width=12cm]{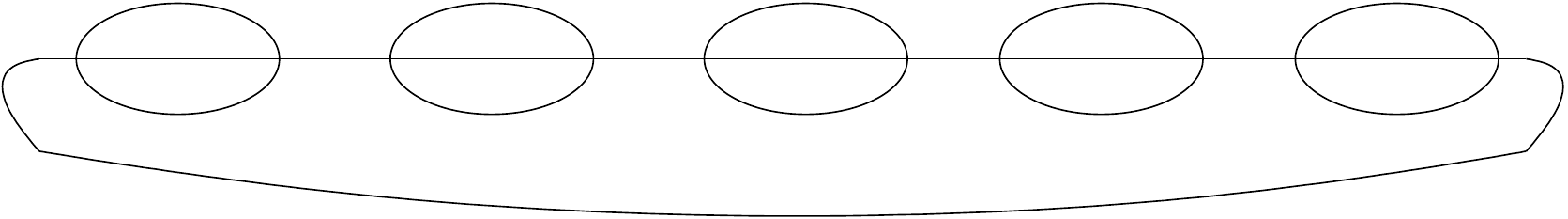}
\caption{
{\small A chain with power counting $2+n/2$}}\label{chain1}
\end{figure}

\begin{figure}
\centering
\centering
\includegraphics[angle=0, width=12cm]{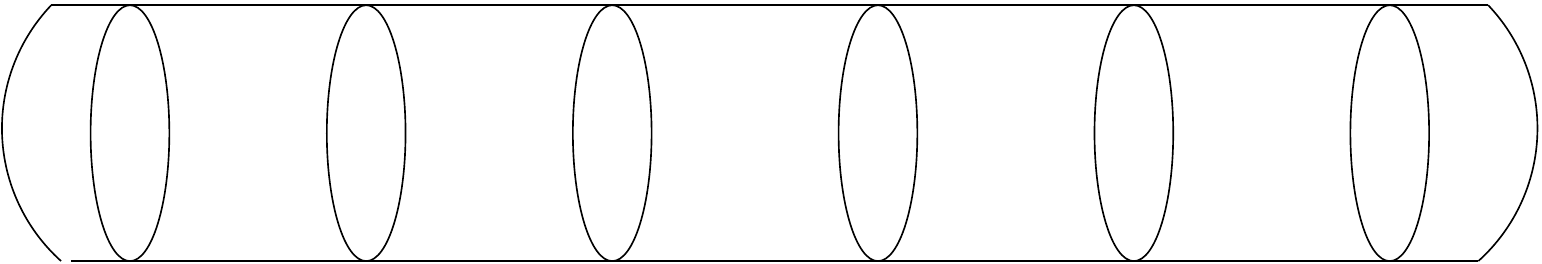}
\caption{
{\small An other chain with power counting $2+n/2$}}
\label{chain2}
\end{figure}

We shall also study in a future publication the behaviour of leading graphs such as the chain of Figures \ref{chain1} and  \ref{chain2}
These are maximal vacuum graphs in the sense that
at order $n$ their linearized power counting is $2+n/2$.
They can be colored easily.

\section*{Acknowledgments}

The work of J.B.G. is supported by the Laboratoire de Physique Th\'eorique d'Orsay
(LPTO, Universit\'e Paris Sud XI) and the Association pour la Promotion Scientifique
de l'Afrique (APSA). J.M. thanks the Laboratoire de Physique Th\'eorique d'Orsay
for their hospitality.

\end{document}